\documentclass[11pt]{amsart}
\usepackage{latexsym,amssymb,amsmath,amscd,amsthm}
\numberwithin{equation}{section}
\theoremstyle{remark}
\newtheorem{theorem}{{\bf THEOREM}}[section]

\newcommand{\bq}{\begin{equation}}
\newcommand{\bea}{\begin{array}}
\newcommand{\eea}{\end{array}}

\newcommand{\ga}{\alpha}
\newcommand{\gep}{\epsilon}
\newcommand{\gD}{\Delta}
\newcommand{\gl}{\lambda}
\newcommand{\gL}{\Lambda}
\newcommand{\gb}{\beta}

\newcommand{\mf}{\mathfrak}

\newcommand{\go}{\omega}
\newcommand{\gO}{\Omega}
\newcommand{\gG}{\Gamma}
\newcommand{\gt}{\theta}
\newcommand{\gs}{\sigma}
\newcommand{\gz}{\zeta}
\newcommand{\gag}{\gamma}
\newcommand{\gd}{\delta}
\newcommand{\pp}{\partial}

\newcommand{\tl}{\tilde}
\newcommand{\na}{\nabla}
\newcommand{\gk}{\kappa}

\newcommand{\bl}{\blacklozenge}

\newcommand{\bgs}{\bigstar}

\newcommand{\gS}{\Sigma}

\newcommand{{\DDD}}{D\!\!\!\!\!\!-}
\title{REMARKS ON THE WDW EQUATION}

\author{Robert Carroll\\University of Illinois, Urbana, IL 61801}

\date{December, 2005\thanks{email: rcarroll@math.uiuc.edu}}

\begin{document}

\bibliographystyle{plain}

\begin{abstract} 
We show a kind of converse to some results of Hall and Reginatto on exact uncertainty
related to the Schr\"odinger and Wheeler-deWitt equations.  Some survey material on
statistical geometrodynamics is also sketched.
\end{abstract}

\maketitle

\tableofcontents

\section{INTRODUCTION}
\renewcommand{\theequation}{1.\arabic{equation}}
\setcounter{equation}{0}

We abbreviate SE for the Schr\"odinger equation and WDW for the Wheeler-deWitt
equation.

\section{EXACT UNCERTAINTY AND THE SE}
\renewcommand{\theequation}{2.\arabic{equation}}
\setcounter{equation}{0}

The exact uncertainty principle of Hall and Reginatto is discussed at length in
\cite{c1,h1,h2,h3,r1}.  Basically following e.g. \cite{h1,h3} one defines
Fisher information via $(\bullet)\,\,F_x=\int dx P(x)[\pp_xlog(P(x))]^2$ and a
Fisher length by $\gd x=F_x^{-1/2}$ where $P(x)$ is a probability density for a
1-D observable x.  The Cramer-Rao inequality says $Var(x)\geq F_x^{-1}$ or simply
$\gD x\geq\gd x$.  For a quantum situation with $P(x)=|\psi(x)|^2$ and $\psi$
satisfying a SE one finds immediatly
\bq\label{1.1}
F_X=\int
dx|\psi|^2\left[\frac{\psi'}{\psi}+\frac{\bar{\psi}'}{\bar{\psi}}\right]^2dx=
\end{equation}
$$=4\int dx\bar{\psi}'\psi'+\int dx|\psi|^2\left[\frac{\psi'}{\psi}-\frac
{\bar{\psi}'}{\bar{\psi}}\right]^2
=\frac{4}{\hbar^2}\left[<p^2>_{\psi}-<p_{cl}^2>_{\psi}\right]$$
where $p_{cl}=(\hbar/2i)[(\psi'/\psi)-(\bar{\psi}'/\bar{\psi})]$ is the classical
momentum observable conjugate to x ($\sim S_X$ for $\psi=Rexp(iS/\hbar)$).
Setting now $p=p_{cl}+p_{nc}$ one obtains after some calculation
$(\bl)\,\,F_x=(4/\hbar^2)(\gD p_{nc})^2=1/(\gd x)^2\Rightarrow \gd
x\gD p_{nc}=\hbar/2$ as a relation between nonclassicality and Fisher
information. Note $<p>_{\psi}=<p_{cl}>_{\psi}$,
$\pp_t|\psi|^2+\pp_x[|\psi|^2m^{-1}p_{cl}]=0$ from the SE, and $(\gD x)(\gD
p)\geq (\gd x)(\gD p)\geq (\gd x)(\gD p_{nc})$.
\\[3mm]\indent
We recall also that from \eqref{1.1} $F_x$ is proportional to the difference of
a quantum and a classical kinetic energy.  Thus $(\hbar^2/4)F_x(1/2m)=(1/2m)
<p^2>_{\psi}-(1/2m)<p^2_{cl}>_{\psi}$ and $E_F=(\hbar^2/8m)F_x$ is added to
$E_{cl}$ to get $E_{quant}$.  By deBroglie-Bohm (dBB) theory there is a quantum
potential
\bq\label{1.2}
Q=\frac{\hbar^2}{8m}\left[\left(\frac{P'}{P}\right)^2-2\frac{P''}{P}\right];\,\,
P=|\psi|^2
\end{equation}
and evidently $(\bgs)\,\,<Q>_{\psi}=\int PQdx=(\hbar^2/8m)F_x$ (upon neglecting
the boundary integral term at $\pm\infty$ - i.e. $P'\to 0$ at $\pm\infty$).
\\[3mm]\indent
Now the exact uncertainty principle (cf. \cite{h1,h3,r1}) looks at momentum
fluctuations $(\clubsuit)\,\,p=\na S+f$ with $<f>=\bar{f}=0$ and replaces a 
classical ensemble energy $<E>_{cl}$ by ($P\sim|\psi|^2$)
\bq\label{1.3}
<E>=\int dx P\left[(2m)^{-1}\overline{|\na S+f|^2}+V\right]=<E>_{cl}+
\int dx P\frac{\overline {f\cdot f}}{2m}
\end{equation}
Upon making an assumption of the form $(\spadesuit)\,\,\overline{f\cdot f}=
\ga(x,P,S,\na P,\na S,\cdots)$ one looks at a modified Hamiltonian
$(\bullet\bullet)\,\,\tl{H}_q[P,S]=\tl{H}_{cl}+\int dx P(\ga/2m)$.  Then,
assuming
\begin{enumerate}
\item
Causality - i.e. $\ga$ depends only on $S,P$ and their first derivatives
\item
Independence for fluctuations of noninteracting uncorrelated ensembles
\item
$f\to L^Tf$ for invertible linear coordinate transformations $x\to L^{-1}x$
\item
Exact uncertainty - i.e. $\ga=\overline{f\cdot f}$ is determined solely by 
uncertainty in position
\end{enumerate}
one arrives at
\bq\label{1.4}
\tl{H}_q=\tl{H}_{cl}+c\int dx\frac{\na P\cdot\na P}{2m P}
\end{equation}
and putting $\hbar=2\sqrt{c}$ with $\psi=\sqrt{P}exp(iS/\hbar)$ a SE is obtained
(cf. Sections 4 and 5 for more detail).
\\[3mm]\indent
As pointed out in \cite{c2} in the SE situation with Q as in \eqref{1.2}, in
3-D one has
\bq\label{1.5}
\int PQd^3x\sim -\frac{\hbar^2}{8m}\int\left[2\gD P-\frac{1}{P}(\na
P)^2\right]d^3x=\frac{\hbar^2}{8m}\int\frac{1}{P}(\na P)^2d^3x
\end{equation}
since $\int_{\gO}\gD Pd^3x=\int_{\pp\gO}\na P\cdot{\bf n}d\gS$ can be assumed zero for
$\na P=0$ on $\pp\gO$. 
Hence (cf. Section 5 for more precision)
\begin{theorem}
Given that any quantum potential for the SE has the form \eqref{1.2} (with
$\na P=0$ on $\pp\gO$) it follows
that the quantization can be identified with momentum fluctuations of the type
studied in \cite{h3} and thus has information content as described by the Fisher
information.  
\end{theorem}

\section{WDW}
\renewcommand{\theequation}{3.\arabic{equation}}
\setcounter{equation}{0}

The same sort of arguments can be applied for the WDW equation (cf. \cite{c2,
h1,h2,p2,r1,s1}).  Thus take an ADM situation
\bq\label{2.1}
ds^2=-(N^2-h^{ij}N_iN_j)+2N_idx^idt+h_{ij}dx^idx^j
\end{equation}
and assume dynamics generated by an action $(\bl\bl)\,\,A=\int dt[\tl{H}+\int
{\mf D}hP\pp_tS]$.  One will have equations of motion $(\bgs\bgs)\,\,
\pp_tP=\gd\tl{H}/\gd S$ and $\pp_tS=-\gd\tl{H}/\gd P$ (cf. \cite{c1,h2}).  A
suitable ``classical" Hamiltonian is 
\bq\label{2.2}
\tl{H}_c[P,S]=
\int{\mf D}hPH_0\left[h_{ij},\frac{\gd S}{\gd h_{ij}}\right];
\end{equation}
$$H_0=
\int dx\left[N\left(\frac{1}{2}G_{ijk\ell}\pi^{ij}\pi^{k\ell}+V(h_{ij})\right)-
2N_i\na_j\pi^{ij}\right]$$
where $G_{ijk\ell}$ is the deWitt (super)metric $(\clubsuit\clubsuit)\,\,
G_{ijk\ell}=(1/\sqrt{h})(h_{ik}h_{j\ell}+h_{i\ell}h_{jk}-h_{ij}h_{k\ell})$ and
$V\sim \hat{c}\sqrt{h}(2\gL-{}^3R)$.  Then thinking of $\pi^{ij}=\gd S/\gd
h_{ij}+ f^{ij}$ and e.g. $\tl{H}_q=\tl{H}_c+(1/2)\int{\mf D}hP\int dx
NG_{ijk\ell}
\overline{f^{ij}f^{k\ell}}$ one arrives via exact uncertainty at a Fisher information
contribution (cf. \cite{f1,f2})
\bq\label{2.3}
\tl{H}_q[P,S]=\tl{H}_{cl}+\frac{c}{2}\int{\mf D}h\int dx
NG_{ijk\ell}\frac{1}{P}\frac {\gd P}{\gd h_{ij}}\frac{\gd P}{\gd h_{k\ell}}
\end{equation}
with $\hbar=2\sqrt{c}$ and $\psi=\sqrt{P}exp(iS/\hbar)$ resulting in
(for $N=1$ and $N_i=0$)
\bq\label{2.4}
\left[-\frac{\hbar^2}{2}\frac{\gd}{\gd h_{ij}}G_{ijk\ell}\frac{\gd}{\gd
h_{k\ell}}+V\right]\psi=0
\end{equation}
with a sandwich ordering ($G_{ijk\ell}$ in the middle).  In general there are
also constraints 
\bq\label{2.5}
\frac{\gd \psi}{\gd N}=\frac{\gd \psi}{\gd
N_i}=\pp_t\psi=0;\,\,\na_j\left(\frac{\gd \psi}{\gd h_{ij}}\right)=0
\end{equation}
We note here (keeping $N=1$ with $N_i=0$)
\bq\label{2.6}
\frac{\gd}{\gd h_{ij}}\left(G_{ijk\ell}\frac{\gd}{\gd
h_{k\ell}}\sqrt{P}e^{iS/\hbar}\right)=
\left[\frac{\gd G_{ijk\ell}}{\gd h_{ij}}\left(
\frac{1}{2}P^{-1/2}\frac{\gd P}{\gd h_{k\ell}}+\frac{iP^{1/2}}{\hbar}
\frac{\gd S}{\gd h_{k\ell}}\right)+\right.
\end{equation}
$$+G_{ijk\ell}\left\{-\frac{1}{4}P^{-3/2}\frac{\gd P}{\gd h_{k\ell}}\frac
{\gd P}{\gd h_{ij}}+\frac{1}{2}P^{-1/2}\frac{\gd^2P}{\gd h_{k\ell}\gd h_{ij}}-
\frac{P^{1/2}}{\hbar^2}\frac{\gd S}{\gd h_{k\ell}}\frac{\gd S}{\gd
h_{ij}}+\right.$$
$$\left.\left.+\frac{i}{2\hbar}P^{-1/2}\left(\frac{\gd P}{\gd
h_{k\ell}}\frac{\gd S}{\gd h_{ij}}+\frac{\gd S}{\gd h_{k\ell}}\frac{\gd P}{\gd
h_{ij}}\right) +\frac{iP^{1/2}}{\hbar}\frac{\gd^2S}{\gd h_{k\ell}\gd
h_{ij}}\right\}\right]e^{iS/\hbar}$$
Therefore writing out the WDW equation gives
\bq\label{2.7}
-\frac{\hbar^2}{4P}\frac{\gd}{\gd h_{ij}}\left[G_{ijk\ell}\frac{\gd P}{\gd
h_{k\ell}}\right]+
\end{equation}
$$+\frac{\hbar^2}{8P^2}G_{ijk\ell}\frac{\gd P}{\gd h_{k\ell}}
\frac{\gd P}{\gd h_{ij}}+G_{ijk\ell}\left[
\frac{\hbar^2}{8P}\frac{\gd^2P}
{\gd h_{ij}\gd h_{ij}}+\frac{1}{2}\frac{\gd S}{\gd h_{k\ell}}
\frac{\gd S}{\gd h_{ij}}\right]+V=0;$$
$$2P\frac{\gd G}{\gd h_{ij}}\frac{\gd S}{\gd
h_{k\ell}}+G\left(\frac{\gd P}{\gd h_{k\ell}}\frac{\gd S}{\gd h_{ij}}+
\frac{\gd S}{\gd h_{k\ell}}\frac{\gd P}{\gd h_{ij}}\right)+2PG\frac
{\gd^2S}{\gd h_{k\ell}\gd h_{ij}}=0$$
\indent
It is useful here to compare with $-(\hbar^2/2m)\psi''+V\psi=0$ which for
$\psi=Rexp(iS/\hbar)$ yields
\bq\label{2.8}
\frac{1}{2m}S_x^2+V+Q=0;\,\,Q=-\frac{\hbar^2}{4m}\frac{R''}{R}=\frac{\hbar^2}{8m}
\left[\frac{2P''}{P}-\left(\frac{P'}{P}\right)^2\right]
\end{equation}
along with $\pp(R^2S')=\pp(PS')=0$ (leading to (2.5)).  The analogues here are then in
particular
\bq\label{2.9}
\frac{1}{2m}S_x^2\sim \frac{1}{2}G_{ijk\ell}\frac{\gd S}{\gd h_{k\ell}}\frac{\gd
S}{\gd
h_{ij}};\,\,Q=\frac{\hbar^2}{8m}\left[\frac{2P''}{P}-\left(\frac{P'}{P}\right)^2
\right]\sim
\end{equation}
$$\sim-\frac{\hbar^2}{4P}\frac{\gd}{\gd h_{ij}}\left[G_{ijk\ell}\frac{\gd P}{\gd
h_{k\ell}}\right]+G_{ijk\ell}\left\{\frac{\hbar^2}{8P^2}\frac{\gd P}{\gd
h_{k\ell}}
\frac{\gd P}{\gd h_{ij}}+\frac{\hbar^2}{4P}\frac{\gd^2P}{\gd h_{ij}\gd
h_{k\ell}}\right\}$$
We note that the Q term arises directly from
\bq\label{2.10}
Q=-\frac{\hbar^2}{2}P^{-1/2}\frac{\gd}{\gd h_{ij}}\left(G_{ijk\ell}\frac{\gd P^{1/2}}
{\gd h_{k\ell}}\right)
\end{equation}
and hence
\bq\label{2.11}
\int {\mf D}f\,PQ=-\frac{\hbar^2}{2}\int {\mf D}f P^{1/2}\frac{\gd}{\gd h_{ij}}
\left(G_{ijk\ell}\frac{\gd P^{1/2}}{\gd h_{k\ell}}\right)
\end{equation}
%The first integral could be set to zero as before provided \eqref{2.11} makes sense
%for $\tl{\pp}\gO$ a suitable boundary expression and this is discussed at some length
%below (cf. in particular \cite{d1} for cautionary remarks).  Crudely we note that
%\bq\label{2.12}
%\gD f=\frac{1}{\sqrt{g}}\pp_m(\sqrt{g}g^{mn}\pp_nf);\,\,\int_{\pp\gO}\na f\cdot d{\bf
%S}=\int_{\gO}\gD fdV
%\end{equation}
%in the sense that (cf. \cite{d2})
%\bq\label{2.13}
%\int_{\gO}\frac{1}{\sqrt{g}}\pp_m(\sqrt{g}g^{mn}\pp_nf)(\sqrt{g}d\gO)=\int_{\pp\gO}
%g^{mn}\pp_nf\cdot dS_m
%\end{equation}  
%where $\pp_n\sim\pp/\pp x^n$.  This suggests heuristically
But from $\int {\mf D}f\gd[\,\,\,]=0$ one has (cf. (4.3))
\bq\label{2.12}
\int{\mf D}f P^{1/2}\frac{\gd}{\gd h_{ij}}\left(G_{ijk\ell}\frac{\gd P^{1/2}}{\gd
h_{k\ell}}\right)=-\int{\mf D}f\frac{\gd P^{1/2}}{\gd h_{ij}}G_{ijk\ell}\frac
{\gd P^{1/2}}{\gd h_{k\ell}}
\end{equation}
This suggests heuristically (see Section 4 for more details of proof and Section 5
for more precision)
\begin{theorem}
Given a WDW equation of the form (3.4) with associated quantum potential given via
(3.10) (or (3.9)) it
follows that the quantum potential can be expressed via momentum fluctuations as in 
(3.3) (for $N=1$).
\end{theorem}

\section{SOME FUNCTIONAL CALCULUS}
\renewcommand{\theequation}{4.\arabic{equation}}
\setcounter{equation}{0}

We go here to \cite{c1,h2,h13,n1} and will first sketch the derivation of (3.4)
following \cite{h1,h2} (cf. also \cite{c1}).  The relevant functional calculus goes as
follows.  One defines a functional F of fields $f$ and sets
\bq\label{4.1}
\gd F=F[f+\gd f]-F[f]=\int dx\frac{\gd F}{\gd f_x}\gd f_x
\end{equation}
Here e.g. $dx\sim d^4x$ and in the space of fields there is assumed to be a measure
${\mf D}f$ such that $\int {\mf D}f\equiv\int {\mf D}f'$ for $f'=f+h$ (cf. \cite{b1,h2}).
Then evidently $(\spadesuit\spadesuit)\,\,\int {\mf D}f(\gd F/\gd f)=0$ when $\int
{\mf D}f\,F[f]<\infty$.  Indeed 
\bq\label{4.2}
0=\int {\mf D}f(F[f+\gd f]-F[f])=\int dx\gd f_x\left(\int{\mf D}f\frac{\gd F}{\gd
f_x}\right)
\end{equation}
and this provides an integration by parts formula
\bq\label{4.3}
\int {\mf D}f\,P\left(\frac{\gd F}{\gd f}\right)=-\int {\mf D}f\,\left(\frac{\gd P}{\gd
f}\right)F
\end{equation}
for $P[f]$ a probability density functional.  Classically a probability density functional
arises in discussing an ensemble of fields and conservation of probability requires
\bq\label{4.4}
\pp_tP+\sum_a\int dx\frac{\gd}{\gd f_x^a}\left.\left(P\frac{\gd H}{\gd g_x^a}\right|_
{g=\gd S/\gd f}\right)
\end{equation}
where $g_x^a$ is the momentum corresponding to $f_x^a$ and one assumes a motion equation
\bq\label{4.5}
\pp_tS+H\left(f,\frac{\gd S}{\gd f},t\right)=0
\end{equation}
The equations of motion here are then
\bq\label{4.6}
\pp_tP=\frac{\gD\tl{H}}{\gD S};\,\,\pp_tS=-\frac{\tl{H}}{\gD P}
\end{equation}
where $(\bullet\bullet\bullet)\,\,\tl{H}(P,S,t)=<H>=\int{\mf D}f PH(f,(\gd S/\gd f),t)$.
The variational theory here involves functionals $I[F]=\int {\mf D} f\,\xi(F,\gd F/\gd f)$
and one can write
\bq\label{4.7}
\gD I=I[F+\gD F]-I[F]=\int{\mf D}f\left[\frac{\pp\xi}{\pp F}\gD F+\int dx\left(\frac
{\pp\xi}{\pp(\gd F/\gd f_x)}\right)\frac{\gd (\gD F)}{\gd f_x}\right]=
\end{equation}
$$=\int{\mf D} f\left[\frac{\pp\xi}{\pp F}-\int dx\frac{\gd}{\gd f_x}\left(\frac
{\pp\xi}{\pp(\gd F/\gd f_x)}\right)\right]\gD F+$$
$$+\int dx\int{\mf D}f\frac{\gd}{\gd f_x}\left[\left(\frac{\pp\xi}{\pp(\gd F/\gd
f_x}\right)\gd F\right]$$
Assuming the term $\int{\mf D}f[\,\,\,]\gD F$ is finite the last integral vanishes and
one obtains $(\bl\bl\bl)\,\,\gD I=\int {\mf D}f(\gD I/\gD F)\gD F$, thus defining a
variational derivative
\bq\label{4.8}
\frac{\gD I}{\gD F}=\frac{\pp\xi}{\pp F}-\int dx\frac{\gd}{\gd f_x}\left(\frac
{\pp\xi}{\pp(\gd F/\gd f_x)}\right)
\end{equation}
In the Hamiltonian theory one can work with a generating function S such that
$(\bgs\bgs\bgs)\,\,g=\gd S/\gd f$ and $\pp_tS+H(f,\gd S/\gd f,t)=0$ (HJ equation) and
solving this is equivalent to $\pp_tf=\gd H/\gd g$ and $\pp_tg=-\gd H/\gd f$ (cf. 
\cite{h2}).  Once S is specified the momentum density $g$ is determinied via 
$g=\gd S/\gd f$ and an ensemble of fields is specified by a probability density functional
$P[f]$ (and not by a phase space density functional$\rho[f.g]$.  In the HJ formulation
one writes $(\clubsuit\clubsuit\clubsuit)\,\,V_x[f]=\pp f_x/\pp t=(\gd H/\gd g)|_{g=
\gd S/\gd f)}$ and hence the associated continuity equation $\pp_t\int {\mf D}fP$ is
\bq\label{4.9}
\pp_tP+\int dx\frac{\gd}{\gd f_x}[PV_x]=0
\end{equation}
provided $<V_x>$ is finite.
\\[3mm]\indent
Now after proving (2.4) one proceeds as follows to produce a SE.  The Hamiltonian
formulation gives $(\spadesuit\spadesuit\spadesuit)\,\,\pp_tP=\gD\tl{H}/\gD S$ and 
$\pp_tS=-\gD\tl{H}/\gd P$ where the ensemble Hamiltonian is
\bq\label{4.10}
\tl{H}=\tl{H}[P,S,t]=<H>=\int{\mf D}f PH[f,\gd S/\gd f,t]
\end{equation}
where P and S are conjugate variables.  The equations $(\spadesuit\spadesuit\spadesuit)$
arise from $\gD\tl{A}=0$ where $\tl{A}=\int dt[-\tl{H}+\int {\mf D}fS\pp_tP$.  One
specializes here to quadratic Hamiltonian functions
\bq\label{4.11}
H_c[f,g,t]=\sum_{a,b}dx K_x^{ab}[f]g_x^ag_x^b+V[f]
\end{equation}
and to this is added a term as in (2.4) to get $\tl{H}$ (which does not depend on S).
Hence from $(\spadesuit\spadesuit\spadesuit)$ with $\pp_tf_x=\gd H_c/\gd g_x$ one obtains
following (4.9)
\bq\label{4.12}
\pp_tP+\int dx\frac{\gd}{\gd f_x}\left[P\frac{\gd H}{\gd g_x}\right]_{g=\gd S/\gd f}=0
\end{equation}
(cf. 4.8)).  The other term in $\tl{H}$ is simply 
\bq\label{4.13}
(\hbar^2/4)\int {\mf D}f\int PK_x^{ab}(\gd P/\gd f_x^a)(\gd P/\gd f_x^b)(1/P^2)
\end{equation}
and
this provides a contribution to the HJ equation via $\pp_tS=-\gD\tl{H}/\gD P$ which will
have the form
\bq\label{4.14}
Q=-\frac{\hbar^2}{4}P^{-1/2}\int dx\frac{\gd}{\gd f_x^a}\left(K_x^{ab}\frac{\gd
P^{1/2}} {\gd f_x^b}\right)
\end{equation}
corresponding to (3.10).  We note further then from (3.12)
\bq\label{4.15}
Q\sim \frac{\hbar^2}{2}\int dx G_{ijk\ell}\frac{\gd P^{1/2}}{\gd h_{ij}}\frac{\gd P^{1/2}}
{\gd h_{k\ell}}\sim\frac{\hbar^2}{8}\int dx G_{ijk\ell}\frac{1}{P}\frac{\gd P}
{\gd h_{ij}}\frac{\gd P}{\gd h_{k\ell}}
\end{equation}
as in (3.3).  Hence Theorem 3.1 is established under the hypotheses indicated concerning
${\mf D}f$ etc.

\section{ENTROPY AND FISHER INFORMATION}
\renewcommand{\theequation}{5.\arabic{equation}}
\setcounter{equation}{0}

We recall first (cf. \cite{b2,c1,c8}) that the relation between the SE and the quantum
potential (QP) is not 1-1.  The QP Q depends on the wave function $\psi=Rexp(iS/\hbar)$
via $Q=-(\hbar^2/2)(\gD R/R)$ for the SE and thus the solution of a quantum HJ equation,
involving S and R(via Q), requires the companion ``continuity" equation to determine
S and R (and thence $\psi$).  There is some lack of uniqueness since Q determines R only
up to uniqueness for solutions of $\gD R+(2m/\hbar^2)QR=0$ and even then the HJ equation
$S_t+\cdots=0$ could introduce still another arbitrary function (cf. \cite{c1,c8}).
Thus to indicate precisely what is said in Theorems 2.1 and 3.1 we rephrase this in the
form
\begin{theorem}
In Theorem 2.1 we see that given a SE described via a probability distribution
$P\,\,(=|\psi|^2)$ one can identify this equation as a quantum model arising from a
classical Hamiltonian $\tl{H}_{cl}$ perturbed by a Fisher information term as in (2.4).
Thus the quantization involves an information content with entropy significance
(cf. here \cite{c2,o1}) for entropy connections).  This suggests that any quantization of
$\tl{H}_{cl}$ arises (or can arise) through momentum perturbations related to Fisher
information and it also suggests that $P=|\psi|^2$ (with $\int Pd^3x=1$) should be 
deemed a requirement for any solution $\psi$ of the related SE (note $\int Pd^3x
=1$ eliminates many putative counterexamples).  Thus once P is specified as a
probability distribution for a wave function $\psi=\sqrt{P}exp(iS/\hbar)$ arising
from a SE corresponding to a quantization of $\tl{H}_{cl}$, then Q can be expressed
via Fisher information.  Similarly given Q as a Fisher information perturbation
of $\tl{H}_{cl}$ (arising from momentum fluctuations involving P as in (2.4)) there is a
unique wave function $\psi=\sqrt{P}exp(iS/\hbar)$ satisfying the corresponding SE.
\end{theorem}
\begin{theorem}
For Theorem 3.1 let us assume there exists a suitable ${\mf D}f$ as in Section 4, which
is a measure in the (super)space of fields $h$.  Then there is an integration by parts
formula (4.3) which removes the need for considering surface terms in 
integrals $\int d^4x$ (cf. \cite{d1} for cautionary remarks about Green's theorem, etc.). 
Consequently given a WDW equation of the form (3.4) with corresponding Q as in (3.10) (and
$\psi=
\sqrt{P}exp(iS/\hbar)$, one can show that the equation can be modelled on a perturbation
of a classical $\tl{H}_c$ via a Fisher information type perturbation as in (3.3)
(cf. here \cite{c1,f1,f2}).  Here P represents a probability density of fields $h_{ij}$
which determine $G_{ijk\ell}$ (and V incidentally) and the very existence of a 
quantum equation (i.e. WDW) seems to require entropy type input via Fisher information
fluctuation of fields.  This suggests that quantum gravity requires a statistical
spacetime (an idea that has appeared before - cf. \cite{c1}).
\end{theorem}
\indent
We sketch now some material from \cite{c3,c17} supporting the idea of a statistical
geometrodynamics (SGD).  Here one builds a model of SGD based on (i) Positing that the
geometry of space is of statistical origin and is explained in terms of the
distinguishability Fisher-Rao (FR) metric and (ii) Assuming the dynamics of the geometry
is derived solely from principles of inference.  There is no external time but an intrinsic
one \`a la \cite{b12}.  A scale factor $\gs(x)$ is required to assign a Riemannian
geometry and it is conjectured that it can be chosen so that the evolving geometry
of space sweeps out a 4-D spacetime.  The procedure defines only a conformal geometry
but that is entirely appropriate d'apr\`es \cite{y2}.  One uses the FR metric in two ways,
one to distinguish neighboring points and the other to distinguish successive states.
Consider then a ``cloud" of dust with coordinate values $y^i\,\,(\,i=1,2,3)$ and estimates
$x^i$ with $p(y|x)dy$ the probability that the particle labeled $x^i$ should have been
labeled $y^i$ (the FR metric encodes the use of probability distributions - instead of
structureless points).  One writes
\bq\label{5.1}
\frac{p(y|x+dx)-p(y|x)}{p(y|x)}=\frac{\pp log[p(y|x)]}{\pp x^i}dx^i
\end{equation}
\bq\label{5.2}
d\gl^2=\int d^4yp(y|x)\frac{\pp log[p(y|x)]}
{\pp x^i}\frac{\pp log[p(y|x)]}{\pp x^j}dx^idx^j=\gag_{ij}dx^idx^j
\end{equation}
and $d\gl^2=0\iff dx^i=0$.  The FR metric $\gag_{ij}$ is the only local Riemannian
metric reflecting the underlying statistical nature of the manifold of distributions
$p(y|x)$ and a scale factor $\gs$ giving a metric $g_{ij}(x)=\gs(x)\gag_{ij}(x)$ is
needed for a Riemannian metric (cf.
\cite{c3,c17}).  Also the metric
$d\gl^2$ is related to the entropy of 
$p(y|x+dx)$ relative to $p(y|x)$, namely
\bq\label{5.3}
S[p(y|x+dx)|p(y|x)]=-\int d^3yp(y|x+dx)log\frac{p(y|x+dx)}{p(y|x)}=-\frac{1}{2}d\gl^2
\end{equation}
and maximizing the relative entropy S is equivalent to minimizing $d\gl^2$.  One thinks
of $d\gl$ as a spatial distance in specifying that the reason that particles at 
$x$ and $x+dx$ are considered close is because they are difficult to distinguish.  To
assign an explicit $p(y|x)$ one assumes the relevant information is given via $<y^i>=x^i$
and the covariance matrix $<(y^i-x^i)(y^j-x^j)>=C^{ij}(x)$; this leads to
\bq\label{5.4}
p(y|x)=\frac{C^{1/2}}{(2\pi)^{3/2}}exp\left[-\frac{1}{2}C_{ij}(y^i-x^i)(y^j-x^j)\right]
\end{equation}
where $C^{ik}C_{kj}=\gd^i_j$ and $C=det(c_{ij})$.  Subsequently to each $x$ one
associates a probability distribution
\bq\label{5.5}
p(y|x,\gag)=\frac{\gag^{1/2}(x)}{(2\pi)^{3/2}}exp\left[-\frac{1}{2}\gag_{ij}
(x)(y^i-x^i)(y^j-x^j)\right]
\end{equation}
where $\gag_{ij}(x)=C_{ij}(x)$ (extreme curvature situations are avoided here).  One
deals with a conformal geometry described via $\gag_{ij}$ and a scale factor $\gs(x)$
will be needed to compare uncertainties at different points; the choice of $\gs$ 
should then be based on making motion ``simple".
\\[3mm]\indent
Thus define a macrostate via
\bq\label{5.6}
P[y|\gag]=\prod_xp(y(x)|x,\gag_{ij}(x))=
\end{equation}
$$=\left[\prod_x\frac{\gag^{1/2}(x)}{(2\pi)^{3/2}}
\right]exp\left[-\frac{1}{2}\sum_x\gag_{ij}(x)(y^i-x^i)(y^j-x^j)\right]$$
Once a dust particle in an earlier state $\gag$ is identified with the label $x$
one assumes that this particle can be assigned the same label $x$ as it evolves into the
later state $\gag+\gD\gag$ (equilocal comoving coordinates).  Then the change between
$P[y|\gag+\gD\gag]$ and $P[y|\gag]$ is denoted by $\gD\ell$ and is measured via their
relative entropy (this is a form of Kullback-Liebler entropy - cf. \cite{c1})
\bq\label{5.7}
S[\gag+\gD\gag|\gag]=-\int\left(\prod_xdy(x)\right)P[y|\gag+\gD\gag]log\frac{P[y|\gag+
\gD\gag]}{P[y|\gag]}=-\frac{1}{2}\gD\ell^2
\end{equation}
Since $P[y|\gag]$ and $P[y|\gag+\gD\gag]$ are products one can write
\bq\label{5.8}
S[\gag+\gD\gag,\gag]=\sum_xS[\gag(x)+\gD\gag(x),\gag(x)]=
\end{equation}
$$=-\frac{1}{2}\sum_x\gD\ell^2(x);
\,\,\gD\ell^2(x)=g^{ijk\ell}\gD\gag_{ij}(x)\gD\gag_{k\ell}(x)$$
where, using \eqref{5.5}
\bq\label{5.9}
g^{ijk\ell}=\int d^3yp(y|x,\gag)\frac{\pp log[p(y|x,\gag)]}{\pp\gag_{ij}}
\frac{\pp log[p(y|x,\gag)]}{\pp \gag_{k\ell}}=
\end{equation}
$$=\frac{1}{4}\left(\gag^{ik}\gag^{ji}+\gag^{i\ell}\gag^{jk}\right)$$
Then $\gD L^2=\sum_x\gD\ell^2(x)$ can be written as an integral if we note that the
density of distinguishable distributions is $\gag^{1/2}$.  Thus the number of
distinguishable distributions, or distinguishable points in the interval $dx$ is
$dx\gag^{1/2}$ ($dx\sim d^3x$) and one has
\bq\label{5.10}
\gD L^2=\int dx\gag^{1/2}\gD\ell^2=\int dx\gag^{1/2}g^{ijk\ell}\gD\gag_{ij}
\gD\gag_{k\ell}
\end{equation}
Thus the effective number of distinguishable points in the interval $dx$ is finite
(due to the intrinsic fuzziness of space).  Now to describe the change $\gD\gag_{ij}(x)$
one introduces an arbitrary time parameter $t$ along a trajectory
\bq\label{5.11}
\gD\gag_{ij}=\gag_{ij}(t+\gD t,x)-\gag_{ij}(t,x)=\pp_t\gag_{ij}\gD t
\end{equation}
Thus $\pp_t\gag_{ij}$ is the ``velocity" of the metric and \eqref{5.10} becomes
\bq\label{5.12}
\gD L^2=\int dx\gag^{1/2}g^{ijk\ell}\pp_t\gag_{ij}\pp_t\gag_{k\ell}\gD t^2
\end{equation}
\\[3mm]\indent
Now go to an arbitrary coordinate frame where equilocal points
at $t$ and $t+\gD t$ have coordinates $x^i$ and $\tl{x}^i=x^i-\gb^i(x)\gD t$.  Then the
metric at $t+\gD t$ transforms into $\tl{\gag}_{ij}$ with
\bq\label{5.13}
\gag_{ij}(t+\gD t,x)=\tl{\gag}_{ij}(t+\gD t,x)-(\na_i\gb_j+\na_j\gb_i)\gD t
\end{equation}
where $\na_i\gb_j=\pp_i\gb_j-\gG^k_{ij}\gb_k$ is the covariant derivative associated to
the metric $\gag_{ij}$.  In the new frame, setting $\tl{\gag}_{ij}(t+\gD t,x)-\gag_{ij}
(t,x)=\gD\gag_{ij}$ one has
\bq\label{5.14}
\gD_{\gb}\gag_{ij}=\gD\gag_{ij}-(\na_i\gb_j+\na_j\gb_i)\gD t\sim
\gD_{\gb}\gag_{ij}=\dot{\gag}_{ij}\gD t
\end{equation}
$$\dot{\gag}_{ij}=\pp_t\gag_{ij}-\na_i\gb_j-\na_j\gb_i$$
leading to
\bq\label{5.15}
\gD_{\gb}L^2=\int dx\gag^{1/2}g^{ijk\ell}\dot{\gag}_{ij}\dot{\gag}_{k\ell}\gD t^2
\end{equation}
\indent
Next one addresses the problem of specifying the best matching criterion, i.e. what
choice of $\gb^i$ provides the best equilocality match.  This is treated as a problem in
inference and asks for minimum $\gD_{\gb}L^2$ over $\gb$.  Hence one gets
\bq\label{5.16}
\gd (\gD_{\gb}L^2)=2\int dx\gag^{1/2}g^{ijk\ell}\dot{\gag}_{ij}\dot{\gag}_{k\ell}\gD t^2
=0\Rightarrow
\end{equation}
$$\Rightarrow \na_{\ell}(2g^{ijk\ell}\dot{\gag}_{ij})=0\equiv
\na_{\ell}\dot{\gag}^{k\ell}=0$$
(using \eqref{5.9} and $\dot{\gag}^{k\ell}=\pp_t\gag^{k\ell}+\na^k\gb^{\ell}+\na^{\ell}
\gb^k$).  These equations determine the shifts $\gb^i$ giving the best matching and
equilocality for the geometry $\gag_{ij}$ and alternatively they could be considered as
constraints on the allowed change $\gD\gag_{ij}=\pp_t\gag_{ij}\gD t$ for given
shifts $\gb^i$.  In describing a putative entropic dynamics one assumes now e.g.
continuous trajectories with each factor in $P[y|\gag]$ evolving continuously through
intermediate states labeled via $\go(x)=\go\gz(x)$ where $\gz(x)$ is a fixed positive
function and $0<\go<\infty$ is a variable parameter (some kind of many fingered time
\`a la Schwinger, Tomonaga, Wheeler, et al).  It is suggested that they dynamics be
determined by an action
\bq\label{5.17}
J=\int_{t_i}^{t_f}dt\int dx\gag^{1/2}[g^{ijk\ell}\dot{\gag}_{ij}\dot{\gag}_{k\ell}]^{1/2}
\end{equation}
The similarities to ``standard" geometrodynamics are striking.

\subsection{INFORMATION DYNAMICS}

We go here to \cite{c5,c6} and consider the idea of introducing some kind of
dynamics in a reasoning process. One looks at the Fisher metric
defined by 
\bq\label{7.39}
g_{\mu\nu}=\int_Xd^4x p_{\gt}(x)\left(\frac{1}{p_{\gt}(x)}\frac{\pp p_{\gt}(x)}{\pp\gt^{\mu}}\right)
\left(\frac{1}{p_{\gt}(x)}\right)\left(\frac{\pp p_{\gt}(x)}{\pp\gt^{\nu}}\right)
\end{equation}
and constructs a Riemannian geometry via
\bq\label{7.40}
\gG_{\gl\nu}^{\gs}=\frac{1}{2}g^{\nu\gs}\left(\frac{\pp g_{\mu\nu}}{\pp\gt^{\gl}}+
\frac{\pp g_{\gl\nu}}{\pp\gt^{\mu}}-\frac{\pp g_{\mu\gl}}{\pp \gt^{\nu}}\right);
\end{equation}
$$R^{\gl}_{\mu\nu\gk}=\frac{\pp\gG^{\gl}_{\mu\nu}}{\pp\gt^{\gk}}-\frac{\pp\gG^{\gl}_{\mu\gk}}
{\pp\gt^{\nu}}+\gG^{\eta}_{\mu\nu}\gG^{\gl}_{\gk\eta}-\gG^{\eta}_{\mu\gk}\gG^{\gl}_{\nu\eta}$$
Then the Ricci tensor is $R_{\mu\gk}=R^{\gl}_{\mu\gl\gk}$ and the curvature scalar is
$R=g^{\mu\gk}R_{\mu\gk}$.  The dynamics associated with this metric can then be described
via functionals
\bq\label{7.41}
J[g_{\mu\nu}]=-\frac{1}{16\pi}\int\sqrt{g(\gt)}R(\gt)d^4\gt
\end{equation}
leading upon variation in $g_{\mu\nu}$ to equations
\bq\label{7.42}
R^{\mu\nu}(\gt) -\frac{1}{2}g^{\mu\nu}(\gt)R(\gt)=0
\end{equation}
Contracting with $g_{\mu\nu}$ gives then the Einstein equations $R^{\mu\nu}(\gt)=0$ (since $R=0$).
J is also invariant under $\gt\to\gt+\gep(\gt)$ and variation here plus contraction leads to a 
contracted Bianchi identity.  Constraints can be built in by adding terms $(1/2)\int\sqrt{g}
T^{\mu\nu}g_{\mu\nu}d^4\gt$ to $J[g_{\mu\nu}]$.  If one is fixed on a given probability distribution
$p(x)$ with variable $\gt^{\mu}$ attached to give $p_{\gt}(x)$ then this could conceivably
describe some gravitational metric based on quantum fluctuations for example.  As examples
a Euclidean metric is produced in 3-space via Gaussian $p(x)$ and complex Gaussians will give
a Lorentz metric in 4-space.

\section{OTHER FORMS OF WDW}
\renewcommand{\theequation}{6.\arabic{equation}}
\setcounter{equation}{0}

In general there are many approaches to WDW and we cite in particular
\cite{a1,a2,a7,a3,b12,b81,c1,c2,c7,
d1,g4,g2,g3,h4,k3,k5,k2,k1,k4,k6,k17,m1,n1,n2,n3,p1,p2,p3,p4,r2,r3,r4,s3,s1,s4,s14,s13,
s2,t1,w1,w2,w3,y1,y2}.
In particular
(for $\phi$ a matter field) the theory of \cite{p3,p4}
leads to a Bohmian form
\bq\label{6.1}
\left\{-\hbar^2\left[\gk G_{ijk\ell}\frac{\gd}{\gd h_{ij}}\frac{\gd}{\gd h_{k\ell}}
+\frac{1}{2}h^{-1/2}\frac{\gd^2}{\gd \phi^2}\right]+V\right\}\psi(h_{ij},\phi)=0;
\end{equation}
$$V=h^{1/2}\left[-\gk^{-1}({}^3R-2\gL)+\frac{1}{2}h^{ij}\pp_i\pp_j\phi+U(\phi)\right]$$
involving (for $A^2\sim P$)
\bq\label{6.2}
\gk G_{ijk\ell}\frac{\gd S}{\gd h_{ij}}\frac{\gd S}{\gd h_{k\ell}}+\frac{1}{2}h^{-1/2}\left(\frac{\gd
S}{\gd \phi}\right)^2+V+Q=0;
\end{equation}
$$Q=-\frac{\hbar^2}{A}\left(\gk G_{ijk\ell}\frac{\gd^2A}{\gd h_{ij}\gd h_{k\ell}}+\frac{h^{-1/2}}{2}\frac
{\gd ^2A}{\gd\phi^2}\right)$$
where the unregularized Q above depends on the regularization and factor ordering
prescribed for the WDW equation.  In addition to (6.2) one has
\bq\label{6.3}
\gk G_{ijk\ell}\frac{\gd}{\gd h_{ij}}\left(A^2\frac{\gd S}{\gd h_{k\ell}}\right)+\frac{h^{-1/2}}{2}
\frac{\gd}{\gd\phi}\left(A^2\frac{\gd S}{\gd\phi}\right)=0
\end{equation}
Other Bohmian situations are indicated in \cite{c1} and we are preparing a survey article.

\end{document}